\documentclass[aps,prl,reprint,floatfix, superscriptaddress]{revtex4-1}

\usepackage[pdftex]{graphicx,color}
\DeclareGraphicsExtensions{.pdf,.png}

\usepackage{grffile}

\usepackage{amsmath, amsthm, amssymb, amsfonts, amsbsy}
\usepackage{mathtools} 
\usepackage{bm}
\usepackage{mathrsfs}  

\usepackage{xcolor}
\definecolor{strawberry}{rgb}{1.0,0.0,0.5}

\definecolor{blueberry}{rgb}{0.015686275,0.2,1.0}

\usepackage[capitalise]{cleveref}
\crefformat{equation}{Eq. (#2#1#3)} 
\Crefformat{equation}{Equation (#2#1#3)} 
\crefrangeformat{equation}{Eqs. (#3#1#4--#5\crefstripprefix{#1}{#2}#6)} 
\Crefrangeformat{equation}{Equations (#3#1#4--#5\crefstripprefix{#1}{#2}#6)} 
\crefmultiformat{equation}{Eqs. (#2#1#3)}{ and~(#2#1#3)}{, (#2#1#3)}{ and~(#2#1#3)} 
\Crefmultiformat{equation}{Equations (#2#1#3)}{ and~(#2#1#3)}{, (#2#1#3)}{ and~(#2#1#3)} 

\usepackage[capitalise]{cleveref}
\crefformat{equation}{Eq. (#2#1#3)} 
\Crefformat{equation}{Equation (#2#1#3)} 
\crefrangeformat{equation}{Eqs. (#3#1#4--#5\crefstripprefix{#1}{#2}#6)} 
\Crefrangeformat{equation}{Equations (#3#1#4--#5\crefstripprefix{#1}{#2}#6)} 
\crefmultiformat{equation}{Eqs. (#2#1#3)}{ and~(#2#1#3)}{, (#2#1#3)}{ and~(#2#1#3)} 
\Crefmultiformat{equation}{Equations (#2#1#3)}{ and~(#2#1#3)}{, (#2#1#3)}{ and~(#2#1#3)} 

\newcommand{\dt}[1]{{d #1 \over d t }}

\newcommand{\pard}[2]{{\partial {#1} \over \partial {#2} }}

\newcommand{\bfr}{{\bf r}}

\newcommand{\bfx}{{\bf x}}

\newcommand{\bfv}{{\bf v}}

\newcommand{\bmG}{{\bm G}}

\newcommand{\bmP}{{\bm P}}
\newcommand{\bmD}{{\bm D}}

\newcommand{\bmA}{{\bm A}}

\newcommand{\bfX}{{\bf X}}

\newcommand{\bGamma}{\bm{\Gamma}}

\newcommand{\bbeta}{{\bm \bbeta}}

\newcommand{\bOmega}{{\bm \Omega}}

\newcommand{\aver}[1]{\left\langle {#1}\right\rangle}

\usepackage{graphicx}
\usepackage{color}

\newcommand{\ds}{\displaystyle}
\newcommand{\beq}{\begin{equation}}
\newcommand{\eeq}{\end{equation}}
\newcommand{\beqa}{\begin{eqnarray}}
\newcommand{\eeqa}{\end{eqnarray}}
\newcommand{\bem}{\begin{math}}
\newcommand{\eem}{\end{math}}
\newcommand{\rar}{{\rightarrow}}

\newcommand{\Rar}{{\Rightarrow}}

\def\strutdepth{\dp\strutbox}
\def\nw#1{\strut\vadjust{\kern-\strutdepth\vtop to0pt{\vss\hbox to\hsize
{\hskip\hsize\hskip5pt$\leftarrow$\hss\strut}}}{\em #1}}
\usepackage{multirow}

\begin{document}

\title{The physics and mathematics of living and dying matter}
\author{Tanniemola B. Liverpool}
\affiliation{School of Mathematics, Fry Building, University of Bristol, BS8 1UG, UK}
\affiliation{The Isaac Newton Institute for Mathematical Sciences, Cambridge CB3 0EH, UK}

\author{Kristian K. M\"uller-Nedebock}
\affiliation{Department of Physics, University of Stellenbosch, Private Bag X1, Matieland 7602 South Africa}
\affiliation{National Institute for Theoretical and Computational Sciences, Stellenbosch, 7600, South Africa}
\affiliation{The Isaac Newton Institute for Mathematical Sciences, Cambridge CB3 0EH, UK}

\author{Xichen Chao}
\affiliation{School of Mathematics, Fry Building, University of Bristol, BS8 1UG, UK}

\author{Chang Yuan}
\affiliation{School of Mathematics, Fry Building, University of Bristol, BS8 1UG, UK}

\date{\today}

\begin{abstract}

We introduce and study a class of active matter models in which we keep track of fuel (stored energy) consumption.
They are by construction, thermodynamically consistent. Using these models it is possible for us to observe and follow how active behaviour develops and also how it dissipates as the energy runs out.  It is also straightforward to define, calculate and keep track of macroscopic thermodynamic quantities.

\end{abstract}

\maketitle

Active matter describes a new class of materials that are composed of elements driven out of equilibrium by internal sources of energy. 
{They promise a novel route to}
functionality in materials design for numerous applications, from drug delivery to metamaterials~\cite{baylis2015self,di2010bacterial, krishnamurthy2016micrometre,stenhammar2016light, frangipane2018dynamic, arlt2018painting,chen2021realization}. 
Early studies of these active systems were motivated by 
biological processes on a wide array of length scales
(from e.g. fluctuations of membranes \cite{Prost_1996}
to flocking birds \cite{PhysRevLett.75.1226,TonerTu95,TONER2005170}), but has increasingly found application in synthetic man-made systems~\cite{Howse2007a,Bricard2013}. 
For obvious pragmatic reasons, commonly studied theoretical models of active matter systems typically
\cite{Julicher2007,Ramaswamy2010,Marchetti2013,Bechinger2016,Gompper_2020}, never reach an
equilibrium state. This  corresponds to infinite reservoirs of the internal (bulk) energy sources that drive them. 
Since this is impossible for any realistic experimental system, it is worth exploring going beyond 
the limitations of this natural simplification.
%
%
In this letter, we study active matter models in which this consumption of stored energy is explicit and the amount of stored energy is {\em finite};  surprisingly we find that by keeping track of this stored energy (or fuel) and being careful about how one deals with fluctuations, a number of conceptual simplifications emerge which allow us  to obtain new insights about their macroscopic behaviour. 

We will use as a paradigmatic example, a two-dimensional flocking model which spontaneously breaks rotational symmetry that can be thought of as a generalisation of the celebrated Vicsek model~\cite{Vicsek1995,Chate2008}. 
%
We have 
$N$ active particles moving in the plane characterised by 
equations of motion 
for their positions on the plane, ${\bf r}_i (t) = \left( x_i,y_i\right)$,  
and their orientations $\theta_i (t)$, $i \in [1, \ldots, N]$ which determine their directions of self-propulsion. We call these  {\em mechanical} variables.
The mechanical variables interact via an aligning potential, $U= -  \sum_{i,j\neq i}^{r_{ij} \le R} J_0 \cos  \left[ \theta_i - \theta_j \right] $, of range $r_{ij} \equiv  |\bfr_i-\bfr_j|=R$.

\begin{figure}[htbp]
    \centering
    \includegraphics[width=0.98\linewidth]{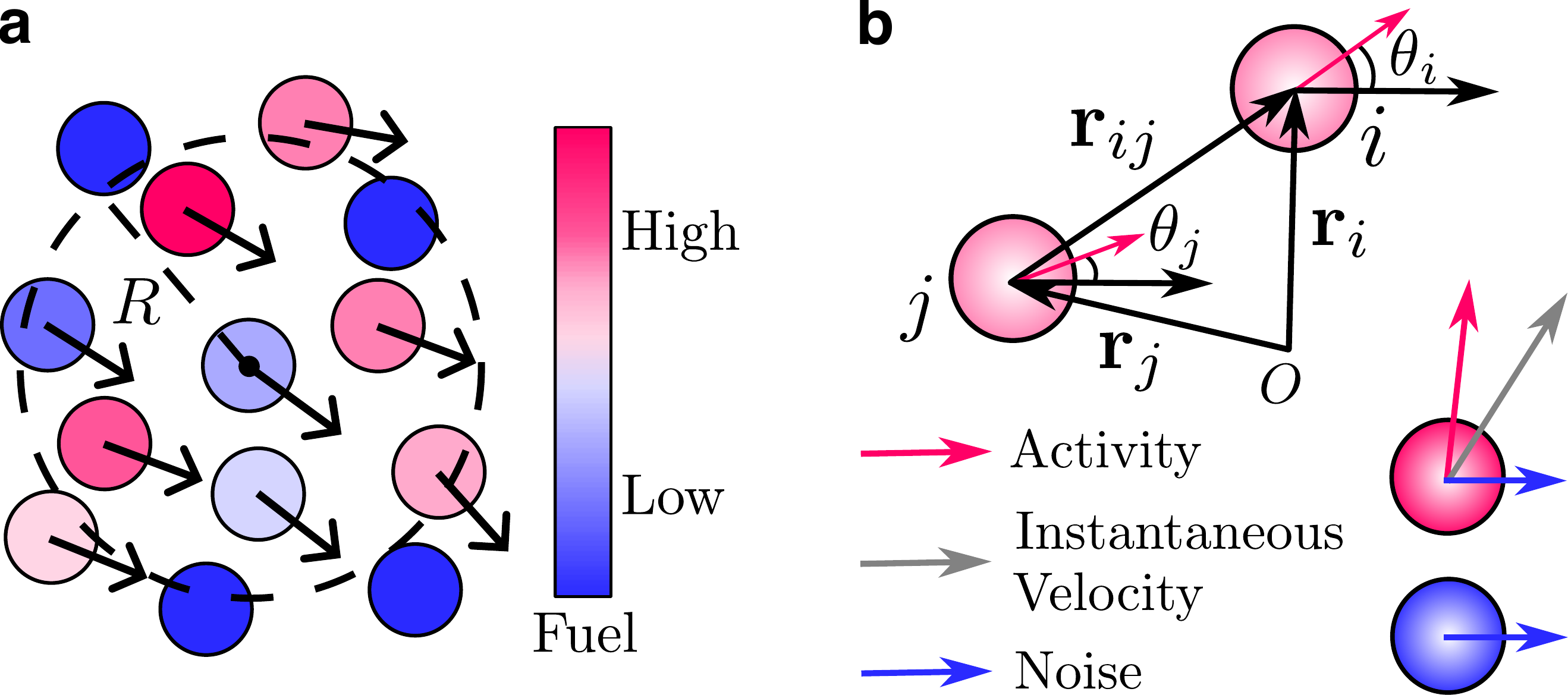}
     \caption{\textbf{Schematic of flocking model with fuel.} \textbf{a}, The color of a particle indicates  the amount of its fuel remaining. Particles that are consuming fuel  self-propel  which we indicate by arrows. $R$ is the aligning interaction radius. \textbf{b}, $\mathbf{r}_{ij}\equiv \mathbf{r}_i - \mathbf{r}_j$ is the separation vector, directed from particle $j$ to $i$. $O$ indicates the origin. $\theta_i$ and $\theta_j$ are the angles between self-propelling directions and the positive $x$ direction, for particle $i$ and $j$ respectively. 
 The instantaneous direction of motion of a particle is dominated by self-propulsion (red arrows) for particles with high fuel consumption rate that are strongly `active', whereas particles with slow fuel consumption rates (e.g. when the fuel has almost run out) are dominated by thermal noise.} 
    \label{Fig1}
\end{figure}

We now define new variables (equivalent to the difference between fuel and products) to keep track of  the  `reaction' or energy consumption that is driving the active motion. 
We will in the following refer to them as {\em stored energy} or {\em fuel} variables. 
We want to construct models in which (1) the active terms are proportional to the {\em rate} of consumption of fuel (stored energy) and vanish when the fuel runs out and (2) when the fuel runs out the system behaves as a passive system with potential energy $U$  that  evolves to a state of equilibrium at the same temperature as its environment with a detailed balance condition.
In this example, it is most natural to have individual fuel variables for each particle,  $n_i (t)$, and at first we will only consider the case in which the consumption of stored energy affects the mechanical variables, generating the active terms, but not vice-versa~\footnote{It is not difficult to generalise the model to do so}. This is equivalent to assuming a very anisotropic  `reaction'  landscape driving the active motion in which the backward reaction is very unlikely. We also define a fuel potential $\Phi= \sum_{i=1}^N \phi  (n_i)$ to keep track of the amount of stored energy within each particle. The function $\phi (n)$ has a minimum at $n=0$ and we initialise our system at time $t=0$ with $n_i (0) = n_0 > 0$, i.e. with a finite amount of fuel.
The active (self-propulsion) terms are proportional to the rate of consumption of fuel giving rise to the stochastic equations of motion
\begin{align}
    \dt{{\mathbf{r}}_i}&= -v_0 \hat{\bm{u}}_i  \dt{{n}_i} - \frac{1}{\zeta_r} {\partial U \over \partial \bfr_i}   + \tilde{\bm \eta}_i^r , \label{dynamics f1}\\
    \dt{{\theta}_i}&=- \frac{1}{\zeta_\theta} {\partial U \over \partial \theta_i}   + \tilde\eta_i^\theta,\label{dynamics f2} \\
    \dt{{n}_i}&= - \frac{1}{\zeta_n} {\partial \Phi \over \partial n_i}   + \tilde\eta_i^n , \label{dynamics f3}
\end{align}
where $\hat{\bm u}_i= (\cos \theta_i, \sin \theta_i)$ and  the fluctuations are encoded in noise terms, $\eta_i^a$.
We recover a Vicsek-like model by taking the limit $\dot n_i=$ constant.

Defining the variables,  
$\bfX_i = (X_i^1, X_i^2, X_i^3 , X_i^4) \equiv (x_i,y_i,\theta_i,n_i)$, and the function $H (\{ \bfX_i \}) \equiv U+\Phi$;
the equations of motion  can be written in the form 
\beq
\bmG_{i}^{-1} \cdot \dt{\bfX_i} = - \pard{H}{\bfX_i} + \pmb{ \tilde\eta}_i \quad  \Rar \quad  \dt{\bfX_i} = - \bmG_{i} \cdot  \pard{H}{\bfX_i} + \pmb{\eta}_{i} \; , 
\label{eq:lang1}
\eeq
defining  the invertible  {\em non-symmetric}  matrix, 
\beq
\bmG_{i}^{-1} (\bfX_i)  = \left( \begin{array}{cccc} \zeta_r & 0 & 0 &  \zeta_r v_0 \cos \theta_i \\ 0 &  \zeta_r & 0 &  \zeta_r v_0 \sin \theta_i \\ 0 & 0 & \zeta_\theta & 0 \\ 0 & 0 & 0 & \zeta_n \end{array} \right)
\eeq
and $\pmb{\eta}_i = \bmG_i \cdot \pmb{\tilde \eta}_i$. So far we have not specified the noise correlations which can have 
\beq
\aver{\eta_{i \alpha}}=0 \quad , \quad \aver{ \eta_{i \alpha} (t)  \eta_{j\beta} (t')  } =  {A}_{i \alpha \beta} \left ( \{ \bfX_k \} \right ) \delta_{ij} \delta (t-t')
\eeq
where ${\bmA}_i$ is a strictly positive definite symmetric matrix. 
If we split $\bmG_i (\bfX_i) $ into symmetric, $\bm{\Gamma}_i (\bfX_i) = \frac12 \left( \bmG_i + \bmG_i^T\right)$ and antisymmetric,  $\bm{\Omega}_i  (\bfX_i)= \frac12 \left( \bmG_i - \bmG_i^T\right)$  parts, $\bmG_i = \bm{\Gamma}_i + \bm{\Omega}_i$, then a useful choice is 
\beq
\bmA_i = 2 \Theta \bGamma_i (\bfX_i) 
\eeq
as long as $\bGamma_i$ remains positive definite.
Because the noise is multiplicative the choice of interpretation (discretisation) of noise and subsequent evaluation of the stochastic integral matters. Two common choices are the It\^o (evaluating the integral at the lower limit) and Stratonovich (evaluating at its mid-point). For the problems we will study here, one can map one discretisation to another (at the level of expectation values) by adding deterministic drift terms~\cite{vankampen,Risken} to the equations of motion~\footnote{For modelling the dynamics of physical problems at equilibrium, one often chooses the discretisation that gives rise to the correct steady state equilibrium probability distribution.}. For the It\^o-scheme, noting that ${\partial \over \partial \bfX_i} \cdot \bGamma_i = {\partial \over \partial \bfX_i} \cdot \bOmega_i = 0$, the Langevin equation (\ref{eq:lang1}) is equivalent to a Fokker-Planck equation for the probability of finding the system at $\{\bfX_i\}$ at time $t$, given an initial condition $\{\bfx_i\}$ at time $0$, $P(\{\bfX_i \}, t | \{\bfx_i\}, 0)$  :
\beq
\pard{P}{t} =  \sum_{i} \pard{}{\bfX_i} \cdot \left[ \bGamma_{i} \cdot  \pard{H}{\bfX_i}  P  + \Theta  \, \bGamma_{i} \cdot \pard{P}{\bfX_i} + \bOmega_{i}  \cdot \pard{H}{\bfX_i}  P \right] \quad , 
\eeq
which has a stable steady state~\cite{Markowich2000,Guionnet2002,Liverpool2020,Cameron2023} given by a Gibbs-Boltzmann distribution with $H$ playing the role of the Hamiltonian: 
\bem
\ds
P_{eq} (\{\bfX_i\})  = \frac1Z e^{- H  (\{\bfX_i \})  / \Theta } \; . \; 
\eem
So our choice for the noise fluctuations leads us to a simple, explicit expression for the steady-state distribution. 

\begin{figure}[htpb!]
    \centering
    \includegraphics[width=0.98\linewidth]{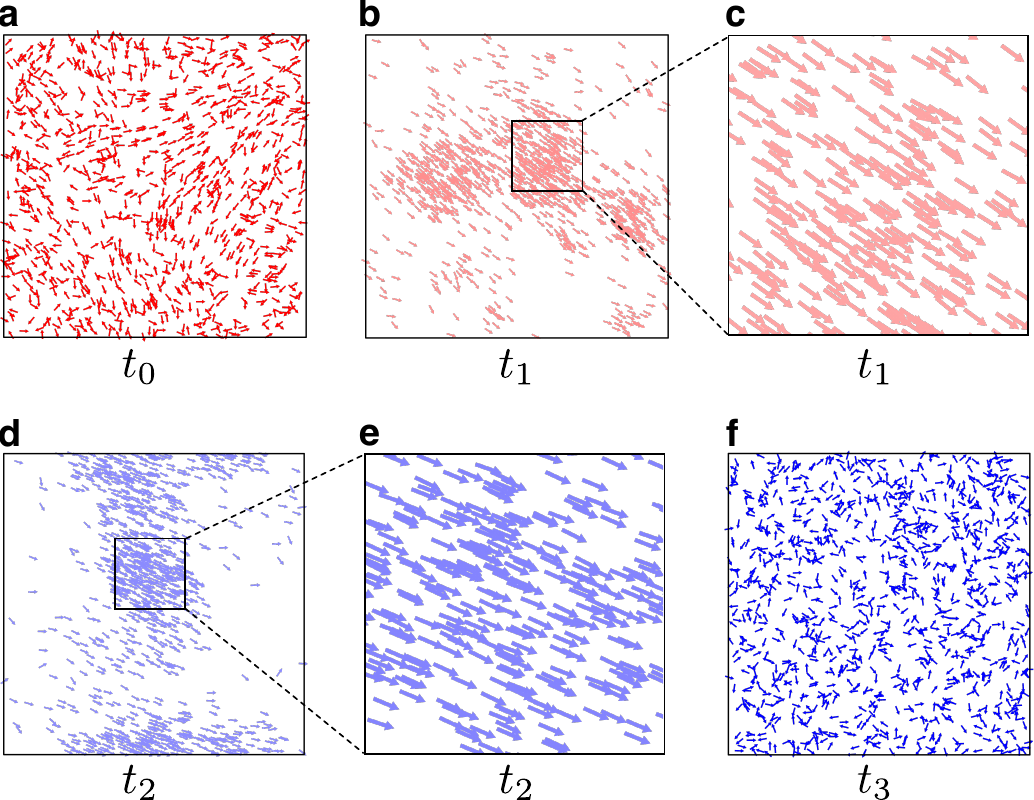}
     \caption{\textbf{Flocking with fuel.} Snapshots from a single trajectory of 1000 particles. Arrows show the instantaneous direction of particles, and colors indicate the amount of fuel remaining. Red (blue) correspond to  high (low)  fuel. \textbf{a}, The simulation starts with random particle positions and directions at $t_0$. \textbf{b-e}, After some time $t_1, t_2 \ (t_1 < t_2)$ before  fuel runs out, fuel consumption gives rise to self-propulsion. The system becomes active and flocks appear. \textbf{f}, After $t_3$, fuel has run out. Particles lose self-propulsion and become passive. The flocking behavior disappears and the passive particles gradually disperse due to thermal noise.       }
    \label{Fig2}
\end{figure}

\protect
\begin{figure*}[htbp]
    \centering
    \includegraphics[width=0.98\linewidth]{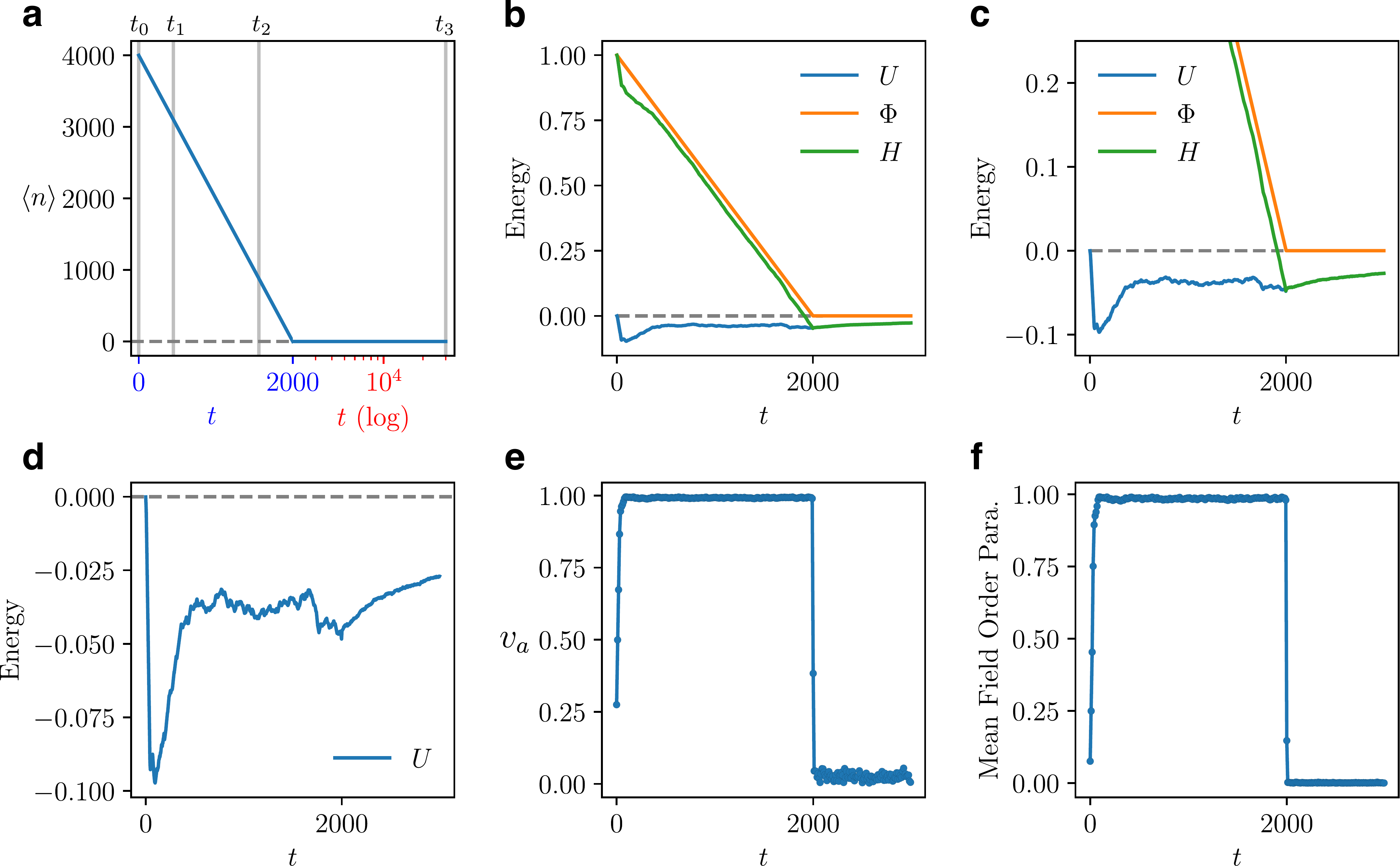}
     \caption{\textbf{The dynamic transition to  active flocking and the transition back to equilibrium.} System size is $L \times L$. $L=15.81$. Number of particles $N=1000$. Number density $\rho\equiv N/L^2=4$. \textbf{a}, Average fuel per  particle $\langle n \rangle$. Standard deviations at all time are less than the line plot width. Fuel runs out at around $t=2000$. Linear time scale  before $t=2000$, logarithmic timescale afterwards. $t_0,t_1,t_2,t_3$ are the times shown in Fig. \ref{Fig2}. \textbf{b-d}, Average  of the rescaled `effective' Hamiltonian $H$, mechanical potential $U$ and fuel potential $\Phi$ (Rescaled by the initial fuel potential $\Phi(n_0)$). \textbf{e}, Absolute value of the average normalized instantaneous velocity $v_a$. 
     See SI for details. \textbf{f}, Mean field order parameter $v_m$. See SI for details.}
    \label{Fig3}
\end{figure*}

This system should show non-equilibrium behaviour as long as the fuel is present and should have equilibrium behaviour in the absence of fuel. 
We verify this by performing numerical simulations. 
We start with the system in an inert passive state (i.e. `dead'), supply the fuel and observe it become active (i.e. `alive') and then watch as the fuel runs out that the system returns to a passive (dead) state. The active state is than by definition a, possibly long-lived but, transient state between two passive states. Depending on the form of the fuel potential $\Phi$, the system can access a pseudo-steady state for which the {\em mechanical} variables are statistically constant. The fuel variables are of course constantly being consumed.

We perform simulations using the  first order Euler-Maruyama algorithm  to integrate  the overdamped Langevin equations~\cite{Higham2001}. We use a two-dimensional square simulation cell with periodic boundary conditions applied in both directions. We initialise our system with particles having random positions and orientations. Particles start with the same amount of fuel initially. The interaction radius $R$ is used as the unit of length, while the time unit is 100 times the time interval between two updates of particles position, orientation and fuel, i.e. $100 \Delta t$.

In order to generate the multiplicative noise matrix for the Langevin dynamics, we perform a Jordan decomposition of $\bGamma_i= \bmP_i \cdot \bmD_i \cdot \bmP_i^{-1}$, with the diagonal matrix, $\bmD_i$. Interestingly,  this naturally leads to a condition required for the matrix $\bGamma_i$ to be strictly positive definite (have all positive eigenvalues): 
\bem
v_0^2 < \frac{4}{\zeta_r {\zeta_n}} \quad . \;
\eem
If it is, so is $\bmD_i$ and  we
can define its square root, $\bmD_i^{1/2}$ by taking the square root of each its diagonal entries and hence, we  can define the noise in Eqn. (\ref{eq:lang1}), $\pmb{\eta}_i (t)=\sqrt{2 \Theta}  \bGamma_i^{1/2} \cdot \pmb{\xi} (t)$ where $\bGamma_i^{1/2}= \bmP_i^{-1} \cdot \bmD_i^{1/2} \cdot \bmP_i^{}$ and $\pmb{\xi} = (\xi_1,\xi_2,\xi_3,\xi_4)$ is a white noise with each component having variance 1. See the SI for details.

We characterize the transitions to/from flocking of the system by measuring the time evolution of the average normalized instantaneous velocity $v_a$.
The instantaneous director describes the normalized instantaneous displacement for each particle,
\begin{align}
    \mathbf{v}_i(t) \equiv \frac{\Delta \mathbf{r}_i(t)}{|\Delta \mathbf{r}_i(t)|} \quad , \label{inst director}
\end{align}
where $\Delta \mathbf{r}_i(t) \equiv \mathbf{r}_i(t+\Delta t) - \mathbf{r}_i(t)$ indicates the displacement of particle $i$ at time $t$. We define the instantaneous order parameter $v_a(t)$ as the magnitude of the ensemble average of the instantaneous directors,
\begin{align}
    v_a(t)\equiv \left\langle |\mathbf{v}_i(t)| \right\rangle = \frac{1}{N}\left| \sum_{i=1}^N \mathbf{v}_i(t) \right| \quad .
\end{align}
$N$ is the number of particles in the system. The time gap $\Delta t$ for calculating Eqn. \ref{inst director} is picked so that the time evolution curve of the order parameter $v_a(t)$ converges. 

We find flocking steady states occur after an initial waiting time and $v_a$ remains steady for a  period before decaying to zero upon running out fuel. 
The rotational friction $\zeta_\theta=200$, the translational friction $\zeta=8$ and the fuel friction $\zeta_n=8$. We use a unit aligning interaction radius $R=1$, and the alignment strength  $J_0=8.75$. 

We also calculate the mean field order parameter $v_m(t)$ which measures the global alignment of all particles' instantaneous directors defined in Eqn. \ref{inst director},
\begin{equation}
    v_m(t) \equiv \frac{1}{N^2} \sum_{i,j=1}^N \mathbf{v}_i(t) \cdot \mathbf{v}_j(t) \quad .
\end{equation}
We study a variety of fuel potentials $\phi(n)$ in simulations which all give similar disorder-order-disorder transitions as the fuel is consumed (See Supplementary Information).  
\begin{align}
    \phi_1(n)&= k \ln(\cosh(n)) \quad , \\
    \phi_2(n)&= \frac{1}{2} k   n^2 \quad ,  \nonumber \\
    \phi_3(n)&= k  n\ln2-\frac{1}{2}k\text{Li}_2(-e^{-2n}) \quad , \nonumber \\
    \phi_4(n)&= k  n-k\ln(1+n) \quad . \nonumber 
\end{align}
The figures in this paper are mostly generated with $\phi_1$ which gives rise to a mean rate of consumption : 
 $f(n)=-\frac{\partial \phi}{\partial n} = -k\tanh(n)$ which remains constant as long as $n$ is finite and greater than 1, $n \ge1$, and goes suddenly to zero as $n \rar 0$. 
 This is helpful to compare with steady states of  `standard' active matter as for $n \ge 1$ the fuel consumption and  the self-propulsion per  particle is approximately constant. The mechanical variables achieve a state where many of their properties are statistically constant. One may call that a {\em pseudo-steady} state.

In Figure \ref{Fig3}, we track the fuel (stored energy) per particle as well as the effective' Hamiltonian $H$, mechanical potential $U$ and fuel potential $\Phi$ as well as the instantaneous and mean field order parameters as a function of time. As expected we see a smooth decay of fuel potential to zero as the fuel is used up, indicated by the decay of $\aver{n}$ to zero, since $n=0$ is the stable minimum of $\phi$. The mechanical potential shows interesting behaviour, there is an initial fast evolution of $U$ from its initial value to a negative average value which remains constant until the fuel runs out, corresponding to an aligned state, however with large fluctuations about  this average~\cite{Derrida2007,Jona-Lasinio2023}. Once the fuel runs out it slowly increases to a new value corresponding to a disordered state (equilibrium) but with significantly smaller fluctuations about the average. The flocking order parameters show a fast evolution to non-zero values which remain constant until the fuel runs out after which the decay to disordered values.
 Equivalent quantities for the other fuel potentials can be found in the SI.
For the four fuel potentials $\phi_i,  1\leq i \leq 4$ we study here, the fuel consumption coefficient $k$ is set so that by starting with the same amount of fuel, the system runs out fuel and goes equilibrium at a similar time. We use $k_1=2,k_2=0.0026, k_3=3, k_4=2$, where $k_i$ corresponds to $\phi_i$. 

\begin{figure}[htbp!]
    \centering
    \includegraphics[width=0.98\linewidth]{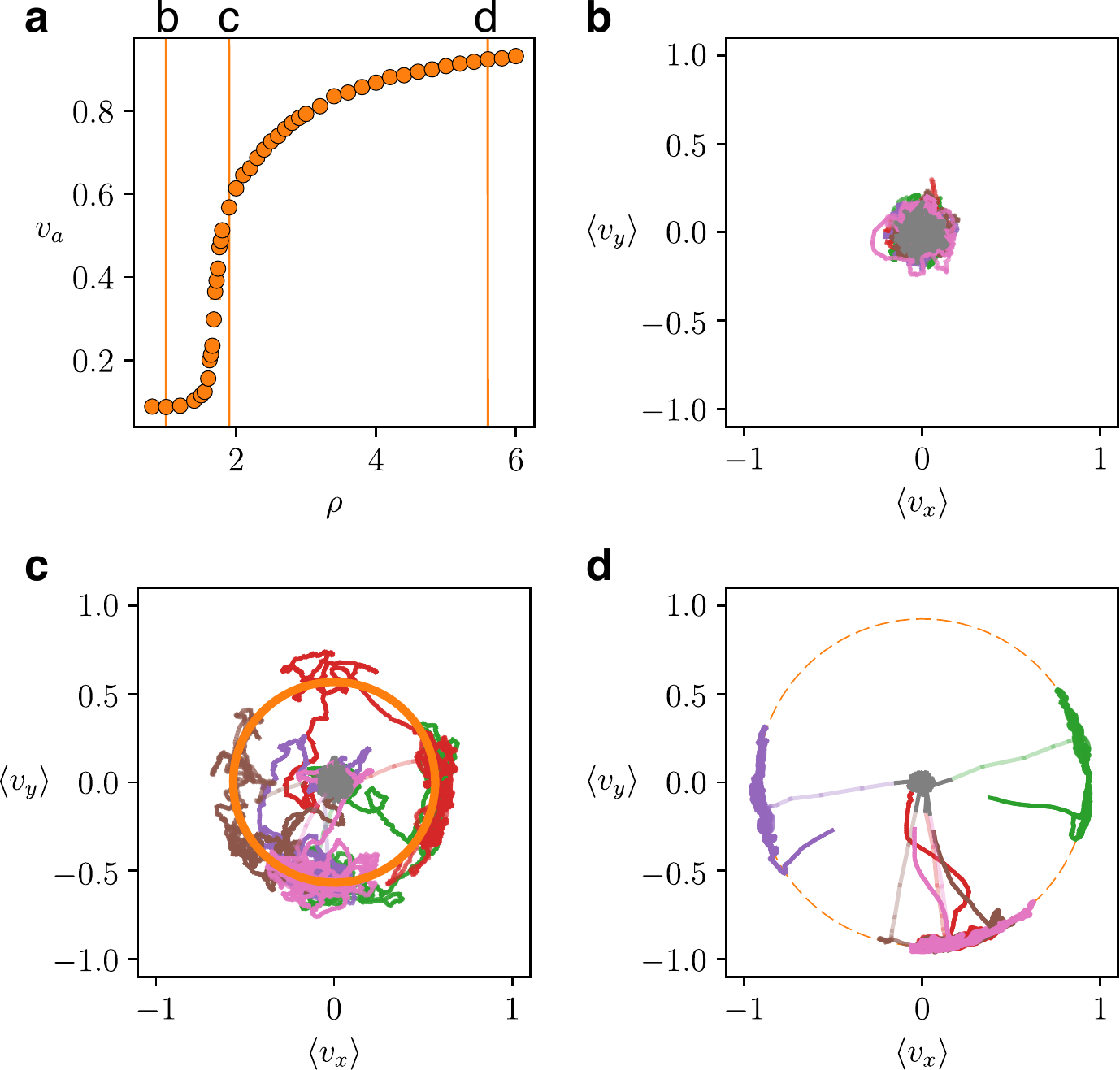}
     \caption{\textbf{Order parameter and phase portraits.} System has  $L=15.81$ and $N=1000$. We calculate the order parameter, $v_a$  using both time and ensemble averages. 
     The time average is made starting when the order parameter attains a steady value and until the fuel runs out. To obtain ensemble averages,  we perform multiple runs with different random seeds for each parameter. See SI for  number of runs for each point. \textbf{a},  Mean order parameter, $v_a$ as a function of the area fraction $\rho$. \textbf{b-d}, Phase portrait of order parameter for simulations with $\rho=1$ (disordered) $\rho=1.9$ (almost ordered) and $\rho=5.6$ (ordered). The $x$ and $y$ axes indicate the averages (sample means) of the $x$ and $y$ components of the normalized instantaneous velocity, $\bfv_i$. Colors show different random seeds. For each trajectory, the color fades as fuel runs out. Once fuel runs out, system is in equilibrium and the color becomes gray.  }
    \label{Fig4}
\end{figure}

To make a link to the Vicsek model~\cite{Vicsek1995}, we consider the behaviour of the system as a function of the  number density: $\rho\equiv N/L^2$. In Figure \ref{Fig4}, we perform multiple runs with different random seeds for different values of $\rho$ (ensemble average). We see a sharp transition from a disordered to a flocking state as a function of $\rho$. Close to the transition point, we observe large fluctuations and the system reaches the steady state slowly, requiring that we average over more trajectories. A similar transition can be observed by increasing the noise amplitudes, e.g. by varying $\Theta$ (see SI). We also compare the individual trajectories of different realisations of the system (starting with different random seeds) to the ensemble average. For each individual system (trajectory), we calculate the average (sample mean) of the instantaneous velocity, $\aver{\bfv_i(t)}=\left( \aver{v_{ix} (t) },\aver{v_{iy} (t)} \right)$ over all the particles at each time $t$. Each seed has a different colour which fades as the fuel runs out. When the fuel is finished, the trajectory becomes grey. 
In the regimes of $\rho$ in which flocking occurs (Figure \ref{Fig4}d), we see that irrespective of initial condition, {\em each} trajectory goes to the circle corresponding to the ensemble average of $v_a$ and stays there until the fuel runs out after which it goes to the centre corresponding to the disordered phase. 
For $\rho$ corresponding to regimes with no flocking (Figure \ref{Fig4}b), the trajectories stay near the centre corresponding to remaining in the disordered phase. At intermediate values of $\rho$ close to the transition, we can observe `partial' flocking (localised near the ensemble average circle but accompanied by large excursions) for each individual trajectory.


In conclusion, we have studied active systems keeping track of the consumption of stored energy (fuel)  that drives the active motion. Using a combination of theory and  simulations, we  identify equilibrium (death) as the true steady state and that the active non-equilibrium state (life) is a possibly long-lived transient between two equilibrium states (death). Furthermore we show that the consumption of fuel is directly linked to non-equilibrium active phenomena we observe and disappears once the stored energy is consumed, using as an example a flocking model that is related to the Vicsek model. 
This framework can be applied to any active model which can be described in terms of Langevin equations (stochastic differential equations).

\begin{acknowledgments}
\paragraph{Acknowledgments} 
The authors would like to thank the Isaac Newton Institute for Mathematical Sciences, Cambridge, for support and hospitality during the programme {\em New statistical physics in living matter}, where part of this work was done. This work was supported by EPSRC grants EP/R014604/1 and EP/T031077/1. XC is supported by Bristol-CSC joint program. 
\end{acknowledgments}

\bibliography{ref_paper,library,bib}

\begin{thebibliography}{31}%
\makeatletter
\providecommand \@ifxundefined [1]{%
 \@ifx{#1\undefined}
}%
\providecommand \@ifnum [1]{%
 \ifnum #1\expandafter \@firstoftwo
 \else \expandafter \@secondoftwo
 \fi
}%
\providecommand \@ifx [1]{%
 \ifx #1\expandafter \@firstoftwo
 \else \expandafter \@secondoftwo
 \fi
}%
\providecommand \natexlab [1]{#1}%
\providecommand \enquote  [1]{``#1''}%
\providecommand \bibnamefont  [1]{#1}%
\providecommand \bibfnamefont [1]{#1}%
\providecommand \citenamefont [1]{#1}%
\providecommand \href@noop [0]{\@secondoftwo}%
\providecommand \href [0]{\begingroup \@sanitize@url \@href}%
\providecommand \@href[1]{\@@startlink{#1}\@@href}%
\providecommand \@@href[1]{\endgroup#1\@@endlink}%
\providecommand \@sanitize@url [0]{\catcode `\\12\catcode `\$12\catcode
  `\&12\catcode `\#12\catcode `\^12\catcode `\_12\catcode `\%12\relax}%
\providecommand \@@startlink[1]{}%
\providecommand \@@endlink[0]{}%
\providecommand \url  [0]{\begingroup\@sanitize@url \@url }%
\providecommand \@url [1]{\endgroup\@href {#1}{\urlprefix }}%
\providecommand \urlprefix  [0]{URL }%
\providecommand \Eprint [0]{\href }%
\providecommand \doibase [0]{http://dx.doi.org/}%
\providecommand \selectlanguage [0]{\@gobble}%
\providecommand \bibinfo  [0]{\@secondoftwo}%
\providecommand \bibfield  [0]{\@secondoftwo}%
\providecommand \translation [1]{[#1]}%
\providecommand \BibitemOpen [0]{}%
\providecommand \bibitemStop [0]{}%
\providecommand \bibitemNoStop [0]{.\EOS\space}%
\providecommand \EOS [0]{\spacefactor3000\relax}%
\providecommand \BibitemShut  [1]{\csname bibitem#1\endcsname}%
\let\auto@bib@innerbib\@empty
\bibitem [{\citenamefont {Baylis}\ \emph {et~al.}(2015)\citenamefont {Baylis},
  \citenamefont {Yeon}, \citenamefont {Thomson}, \citenamefont {Kazerooni},
  \citenamefont {Wang}, \citenamefont {St.~John}, \citenamefont {Lim},
  \citenamefont {Chien}, \citenamefont {Lee}, \citenamefont {Zhang} \emph
  {et~al.}}]{baylis2015self}%
  \BibitemOpen
  \bibfield  {author} {\bibinfo {author} {\bibfnamefont {J.~R.}\ \bibnamefont
  {Baylis}}, \bibinfo {author} {\bibfnamefont {J.~H.}\ \bibnamefont {Yeon}},
  \bibinfo {author} {\bibfnamefont {M.~H.}\ \bibnamefont {Thomson}}, \bibinfo
  {author} {\bibfnamefont {A.}~\bibnamefont {Kazerooni}}, \bibinfo {author}
  {\bibfnamefont {X.}~\bibnamefont {Wang}}, \bibinfo {author} {\bibfnamefont
  {A.~E.}\ \bibnamefont {St.~John}}, \bibinfo {author} {\bibfnamefont {E.~B.}\
  \bibnamefont {Lim}}, \bibinfo {author} {\bibfnamefont {D.}~\bibnamefont
  {Chien}}, \bibinfo {author} {\bibfnamefont {A.}~\bibnamefont {Lee}}, \bibinfo
  {author} {\bibfnamefont {J.~Q.}\ \bibnamefont {Zhang}},  \emph {et~al.},\
  }\href@noop {} {\bibfield  {journal} {\bibinfo  {journal} {Science advances}\
  }\textbf {\bibinfo {volume} {1}},\ \bibinfo {pages} {e1500379} (\bibinfo
  {year} {2015})}\BibitemShut {NoStop}%
\bibitem [{\citenamefont {Di~Leonardo}\ \emph {et~al.}(2010)\citenamefont
  {Di~Leonardo}, \citenamefont {Angelani}, \citenamefont {Dell’Arciprete},
  \citenamefont {Ruocco}, \citenamefont {Iebba}, \citenamefont {Schippa},
  \citenamefont {Conte}, \citenamefont {Mecarini}, \citenamefont {De~Angelis},\
  and\ \citenamefont {Di~Fabrizio}}]{di2010bacterial}%
  \BibitemOpen
  \bibfield  {author} {\bibinfo {author} {\bibfnamefont {R.}~\bibnamefont
  {Di~Leonardo}}, \bibinfo {author} {\bibfnamefont {L.}~\bibnamefont
  {Angelani}}, \bibinfo {author} {\bibfnamefont {D.}~\bibnamefont
  {Dell’Arciprete}}, \bibinfo {author} {\bibfnamefont {G.}~\bibnamefont
  {Ruocco}}, \bibinfo {author} {\bibfnamefont {V.}~\bibnamefont {Iebba}},
  \bibinfo {author} {\bibfnamefont {S.}~\bibnamefont {Schippa}}, \bibinfo
  {author} {\bibfnamefont {M.~P.}\ \bibnamefont {Conte}}, \bibinfo {author}
  {\bibfnamefont {F.}~\bibnamefont {Mecarini}}, \bibinfo {author}
  {\bibfnamefont {F.}~\bibnamefont {De~Angelis}}, \ and\ \bibinfo {author}
  {\bibfnamefont {E.}~\bibnamefont {Di~Fabrizio}},\ }\href@noop {} {\bibfield
  {journal} {\bibinfo  {journal} {Proceedings of the National Academy of
  Sciences}\ }\textbf {\bibinfo {volume} {107}},\ \bibinfo {pages} {9541}
  (\bibinfo {year} {2010})}\BibitemShut {NoStop}%
\bibitem [{\citenamefont {Krishnamurthy}\ \emph {et~al.}(2016)\citenamefont
  {Krishnamurthy}, \citenamefont {Ghosh}, \citenamefont {Chatterji},
  \citenamefont {Ganapathy},\ and\ \citenamefont
  {Sood}}]{krishnamurthy2016micrometre}%
  \BibitemOpen
  \bibfield  {author} {\bibinfo {author} {\bibfnamefont {S.}~\bibnamefont
  {Krishnamurthy}}, \bibinfo {author} {\bibfnamefont {S.}~\bibnamefont
  {Ghosh}}, \bibinfo {author} {\bibfnamefont {D.}~\bibnamefont {Chatterji}},
  \bibinfo {author} {\bibfnamefont {R.}~\bibnamefont {Ganapathy}}, \ and\
  \bibinfo {author} {\bibfnamefont {A.}~\bibnamefont {Sood}},\ }\href@noop {}
  {\bibfield  {journal} {\bibinfo  {journal} {Nature Physics}\ }\textbf
  {\bibinfo {volume} {12}},\ \bibinfo {pages} {1134} (\bibinfo {year}
  {2016})}\BibitemShut {NoStop}%
\bibitem [{\citenamefont {Stenhammar}\ \emph {et~al.}(2016)\citenamefont
  {Stenhammar}, \citenamefont {Wittkowski}, \citenamefont {Marenduzzo},\ and\
  \citenamefont {Cates}}]{stenhammar2016light}%
  \BibitemOpen
  \bibfield  {author} {\bibinfo {author} {\bibfnamefont {J.}~\bibnamefont
  {Stenhammar}}, \bibinfo {author} {\bibfnamefont {R.}~\bibnamefont
  {Wittkowski}}, \bibinfo {author} {\bibfnamefont {D.}~\bibnamefont
  {Marenduzzo}}, \ and\ \bibinfo {author} {\bibfnamefont {M.~E.}\ \bibnamefont
  {Cates}},\ }\href@noop {} {\bibfield  {journal} {\bibinfo  {journal} {Science
  advances}\ }\textbf {\bibinfo {volume} {2}},\ \bibinfo {pages} {e1501850}
  (\bibinfo {year} {2016})}\BibitemShut {NoStop}%
\bibitem [{\citenamefont {Frangipane}\ \emph {et~al.}(2018)\citenamefont
  {Frangipane}, \citenamefont {Dell'Arciprete}, \citenamefont {Petracchini},
  \citenamefont {Maggi}, \citenamefont {Saglimbeni}, \citenamefont {Bianchi},
  \citenamefont {Vizsnyiczai}, \citenamefont {Bernardini},\ and\ \citenamefont
  {Di~Leonardo}}]{frangipane2018dynamic}%
  \BibitemOpen
  \bibfield  {author} {\bibinfo {author} {\bibfnamefont {G.}~\bibnamefont
  {Frangipane}}, \bibinfo {author} {\bibfnamefont {D.}~\bibnamefont
  {Dell'Arciprete}}, \bibinfo {author} {\bibfnamefont {S.}~\bibnamefont
  {Petracchini}}, \bibinfo {author} {\bibfnamefont {C.}~\bibnamefont {Maggi}},
  \bibinfo {author} {\bibfnamefont {F.}~\bibnamefont {Saglimbeni}}, \bibinfo
  {author} {\bibfnamefont {S.}~\bibnamefont {Bianchi}}, \bibinfo {author}
  {\bibfnamefont {G.}~\bibnamefont {Vizsnyiczai}}, \bibinfo {author}
  {\bibfnamefont {M.~L.}\ \bibnamefont {Bernardini}}, \ and\ \bibinfo {author}
  {\bibfnamefont {R.}~\bibnamefont {Di~Leonardo}},\ }\href@noop {} {\bibfield
  {journal} {\bibinfo  {journal} {Elife}\ }\textbf {\bibinfo {volume} {7}},\
  \bibinfo {pages} {e36608} (\bibinfo {year} {2018})}\BibitemShut {NoStop}%
\bibitem [{\citenamefont {Arlt}\ \emph {et~al.}(2018)\citenamefont {Arlt},
  \citenamefont {Martinez}, \citenamefont {Dawson}, \citenamefont {Pilizota},\
  and\ \citenamefont {Poon}}]{arlt2018painting}%
  \BibitemOpen
  \bibfield  {author} {\bibinfo {author} {\bibfnamefont {J.}~\bibnamefont
  {Arlt}}, \bibinfo {author} {\bibfnamefont {V.~A.}\ \bibnamefont {Martinez}},
  \bibinfo {author} {\bibfnamefont {A.}~\bibnamefont {Dawson}}, \bibinfo
  {author} {\bibfnamefont {T.}~\bibnamefont {Pilizota}}, \ and\ \bibinfo
  {author} {\bibfnamefont {W.~C.}\ \bibnamefont {Poon}},\ }\href@noop {}
  {\bibfield  {journal} {\bibinfo  {journal} {Nature communications}\ }\textbf
  {\bibinfo {volume} {9}},\ \bibinfo {pages} {768} (\bibinfo {year}
  {2018})}\BibitemShut {NoStop}%
\bibitem [{\citenamefont {Chen}\ \emph {et~al.}(2021)\citenamefont {Chen},
  \citenamefont {Li}, \citenamefont {Scheibner}, \citenamefont {Vitelli},\ and\
  \citenamefont {Huang}}]{chen2021realization}%
  \BibitemOpen
  \bibfield  {author} {\bibinfo {author} {\bibfnamefont {Y.}~\bibnamefont
  {Chen}}, \bibinfo {author} {\bibfnamefont {X.}~\bibnamefont {Li}}, \bibinfo
  {author} {\bibfnamefont {C.}~\bibnamefont {Scheibner}}, \bibinfo {author}
  {\bibfnamefont {V.}~\bibnamefont {Vitelli}}, \ and\ \bibinfo {author}
  {\bibfnamefont {G.}~\bibnamefont {Huang}},\ }\href@noop {} {\bibfield
  {journal} {\bibinfo  {journal} {Nature communications}\ }\textbf {\bibinfo
  {volume} {12}},\ \bibinfo {pages} {5935} (\bibinfo {year}
  {2021})}\BibitemShut {NoStop}%
\bibitem [{\citenamefont {Prost}\ and\ \citenamefont
  {Bruinsma}(1996)}]{Prost_1996}%
  \BibitemOpen
  \bibfield  {author} {\bibinfo {author} {\bibfnamefont {J.}~\bibnamefont
  {Prost}}\ and\ \bibinfo {author} {\bibfnamefont {R.}~\bibnamefont
  {Bruinsma}},\ }\href {\doibase 10.1209/epl/i1996-00340-1} {\bibfield
  {journal} {\bibinfo  {journal} {Europhysics Letters ({EPL})}\ }\textbf
  {\bibinfo {volume} {33}},\ \bibinfo {pages} {321} (\bibinfo {year}
  {1996})}\BibitemShut {NoStop}%
\bibitem [{\citenamefont {Vicsek}\ \emph
  {et~al.}(1995{\natexlab{a}})\citenamefont {Vicsek}, \citenamefont {Czir\'ok},
  \citenamefont {Ben-Jacob}, \citenamefont {Cohen},\ and\ \citenamefont
  {Shochet}}]{PhysRevLett.75.1226}%
  \BibitemOpen
  \bibfield  {author} {\bibinfo {author} {\bibfnamefont {T.}~\bibnamefont
  {Vicsek}}, \bibinfo {author} {\bibfnamefont {A.}~\bibnamefont {Czir\'ok}},
  \bibinfo {author} {\bibfnamefont {E.}~\bibnamefont {Ben-Jacob}}, \bibinfo
  {author} {\bibfnamefont {I.}~\bibnamefont {Cohen}}, \ and\ \bibinfo {author}
  {\bibfnamefont {O.}~\bibnamefont {Shochet}},\ }\href {\doibase
  10.1103/PhysRevLett.75.1226} {\bibfield  {journal} {\bibinfo  {journal}
  {Phys. Rev. Lett.}\ }\textbf {\bibinfo {volume} {75}},\ \bibinfo {pages}
  {1226} (\bibinfo {year} {1995}{\natexlab{a}})}\BibitemShut {NoStop}%
\bibitem [{\citenamefont {Toner}\ and\ \citenamefont {Tu}(1995)}]{TonerTu95}%
  \BibitemOpen
  \bibfield  {author} {\bibinfo {author} {\bibfnamefont {J.}~\bibnamefont
  {Toner}}\ and\ \bibinfo {author} {\bibfnamefont {Y.}~\bibnamefont {Tu}},\
  }\href@noop {} {\bibfield  {journal} {\bibinfo  {journal} {Phys. Rev. Lett.}\
  }\textbf {\bibinfo {volume} {75}},\ \bibinfo {pages} {4326} (\bibinfo {year}
  {1995})}\BibitemShut {NoStop}%
\bibitem [{\citenamefont {Toner}\ \emph {et~al.}(2005)\citenamefont {Toner},
  \citenamefont {Tu},\ and\ \citenamefont {Ramaswamy}}]{TONER2005170}%
  \BibitemOpen
  \bibfield  {author} {\bibinfo {author} {\bibfnamefont {J.}~\bibnamefont
  {Toner}}, \bibinfo {author} {\bibfnamefont {Y.}~\bibnamefont {Tu}}, \ and\
  \bibinfo {author} {\bibfnamefont {S.}~\bibnamefont {Ramaswamy}},\ }\href
  {\doibase https://doi.org/10.1016/j.aop.2005.04.011} {\bibfield  {journal}
  {\bibinfo  {journal} {Annals of Physics}\ }\textbf {\bibinfo {volume}
  {318}},\ \bibinfo {pages} {170} (\bibinfo {year} {2005})},\ \bibinfo {note}
  {special Issue}\BibitemShut {NoStop}%
\bibitem [{\citenamefont {Howse}\ \emph {et~al.}(2007)\citenamefont {Howse},
  \citenamefont {Jones}, \citenamefont {Ryan}, \citenamefont {Gough},
  \citenamefont {Vafabakhsh},\ and\ \citenamefont {Golestanian}}]{Howse2007a}%
  \BibitemOpen
  \bibfield  {author} {\bibinfo {author} {\bibfnamefont {J.~R.}\ \bibnamefont
  {Howse}}, \bibinfo {author} {\bibfnamefont {R.~a.~L.}\ \bibnamefont {Jones}},
  \bibinfo {author} {\bibfnamefont {A.~J.}\ \bibnamefont {Ryan}}, \bibinfo
  {author} {\bibfnamefont {T.}~\bibnamefont {Gough}}, \bibinfo {author}
  {\bibfnamefont {R.}~\bibnamefont {Vafabakhsh}}, \ and\ \bibinfo {author}
  {\bibfnamefont {R.}~\bibnamefont {Golestanian}},\ }\href {\doibase
  10.1103/PhysRevLett.99.048102} {\bibfield  {journal} {\bibinfo  {journal}
  {Phys. Rev. Lett.}\ }\textbf {\bibinfo {volume} {99}},\ \bibinfo {pages} {8}
  (\bibinfo {year} {2007})},\ \Eprint {http://arxiv.org/abs/0706.4406}
  {arXiv:0706.4406} \BibitemShut {NoStop}%
\bibitem [{\citenamefont {Bricard}\ \emph {et~al.}(2013)\citenamefont
  {Bricard}, \citenamefont {Caussin}, \citenamefont {Desreumaux}, \citenamefont
  {Dauchot},\ and\ \citenamefont {Bartolo}}]{Bricard2013}%
  \BibitemOpen
  \bibfield  {author} {\bibinfo {author} {\bibfnamefont {A.}~\bibnamefont
  {Bricard}}, \bibinfo {author} {\bibfnamefont {J.-B.}\ \bibnamefont
  {Caussin}}, \bibinfo {author} {\bibfnamefont {N.}~\bibnamefont {Desreumaux}},
  \bibinfo {author} {\bibfnamefont {O.}~\bibnamefont {Dauchot}}, \ and\
  \bibinfo {author} {\bibfnamefont {D.}~\bibnamefont {Bartolo}},\ }\href
  {\doibase 10.1038/nature12673} {\bibfield  {journal} {\bibinfo  {journal}
  {Nature}\ }\textbf {\bibinfo {volume} {503}},\ \bibinfo {pages} {95}
  (\bibinfo {year} {2013})}\BibitemShut {NoStop}%
\bibitem [{\citenamefont {Julicher}\ \emph {et~al.}(2007)\citenamefont
  {Julicher}, \citenamefont {Kruse}, \citenamefont {Prost},\ and\ \citenamefont
  {Joanny}}]{Julicher2007}%
  \BibitemOpen
  \bibfield  {author} {\bibinfo {author} {\bibfnamefont {F.}~\bibnamefont
  {Julicher}}, \bibinfo {author} {\bibfnamefont {K.}~\bibnamefont {Kruse}},
  \bibinfo {author} {\bibfnamefont {J.}~\bibnamefont {Prost}}, \ and\ \bibinfo
  {author} {\bibfnamefont {J.}~\bibnamefont {Joanny}},\ }\href {\doibase
  10.1016/j.physrep.2007.02.018} {\bibfield  {journal} {\bibinfo  {journal}
  {Phys. Rep.}\ }\textbf {\bibinfo {volume} {449}},\ \bibinfo {pages} {3}
  (\bibinfo {year} {2007})}\BibitemShut {NoStop}%
\bibitem [{\citenamefont {Ramaswamy}(2010)}]{Ramaswamy2010}%
  \BibitemOpen
  \bibfield  {author} {\bibinfo {author} {\bibfnamefont {S.}~\bibnamefont
  {Ramaswamy}},\ }\href {\doibase 10.1146/annurev-conmatphys-070909-104101}
  {\bibfield  {journal} {\bibinfo  {journal} {Annu. Rev. Condens. Matter
  Phys.}\ }\textbf {\bibinfo {volume} {1}},\ \bibinfo {pages} {323} (\bibinfo
  {year} {2010})}\BibitemShut {NoStop}%
\bibitem [{\citenamefont {Marchetti}\ \emph {et~al.}(2013)\citenamefont
  {Marchetti}, \citenamefont {Joanny}, \citenamefont {Ramaswamy}, \citenamefont
  {Liverpool}, \citenamefont {Prost}, \citenamefont {Rao},\ and\ \citenamefont
  {Simha}}]{Marchetti2013}%
  \BibitemOpen
  \bibfield  {author} {\bibinfo {author} {\bibfnamefont {M.~C.}\ \bibnamefont
  {Marchetti}}, \bibinfo {author} {\bibfnamefont {J.~F.}\ \bibnamefont
  {Joanny}}, \bibinfo {author} {\bibfnamefont {S.}~\bibnamefont {Ramaswamy}},
  \bibinfo {author} {\bibfnamefont {T.~B.}\ \bibnamefont {Liverpool}}, \bibinfo
  {author} {\bibfnamefont {J.}~\bibnamefont {Prost}}, \bibinfo {author}
  {\bibfnamefont {M.}~\bibnamefont {Rao}}, \ and\ \bibinfo {author}
  {\bibfnamefont {R.~A.}\ \bibnamefont {Simha}},\ }\href {\doibase
  10.1103/RevModPhys.85.1143} {\bibfield  {journal} {\bibinfo  {journal} {Rev.
  Mod. Phys.}\ }\textbf {\bibinfo {volume} {85}},\ \bibinfo {pages} {1143}
  (\bibinfo {year} {2013})}\BibitemShut {NoStop}%
\bibitem [{\citenamefont {Bechinger}\ \emph {et~al.}(2016)\citenamefont
  {Bechinger}, \citenamefont {{Di Leonardo}}, \citenamefont {L{\"{o}}wen},
  \citenamefont {Reichhardt}, \citenamefont {Volpe},\ and\ \citenamefont
  {Volpe}}]{Bechinger2016}%
  \BibitemOpen
  \bibfield  {author} {\bibinfo {author} {\bibfnamefont {C.}~\bibnamefont
  {Bechinger}}, \bibinfo {author} {\bibfnamefont {R.}~\bibnamefont {{Di
  Leonardo}}}, \bibinfo {author} {\bibfnamefont {H.}~\bibnamefont
  {L{\"{o}}wen}}, \bibinfo {author} {\bibfnamefont {C.}~\bibnamefont
  {Reichhardt}}, \bibinfo {author} {\bibfnamefont {G.}~\bibnamefont {Volpe}}, \
  and\ \bibinfo {author} {\bibfnamefont {G.}~\bibnamefont {Volpe}},\ }\href
  {\doibase 10.1103/RevModPhys.88.045006} {\bibfield  {journal} {\bibinfo
  {journal} {Rev. Mod. Phys.}\ }\textbf {\bibinfo {volume} {88}} (\bibinfo
  {year} {2016}),\ 10.1103/RevModPhys.88.045006},\ \Eprint
  {http://arxiv.org/abs/1602.00081} {arXiv:1602.00081} \BibitemShut {NoStop}%
\bibitem [{\citenamefont {Gompper}\ \emph {et~al.}(2020)\citenamefont
  {Gompper}, \citenamefont {Winkler}, \citenamefont {Speck}, \citenamefont
  {Solon}, \citenamefont {Nardini}, \citenamefont {Peruani}, \citenamefont
  {Löwen}, \citenamefont {Golestanian}, \citenamefont {Kaupp}, \citenamefont
  {Alvarez}, \citenamefont {Ki{\o}rboe}, \citenamefont {Lauga}, \citenamefont
  {Poon}, \citenamefont {DeSimone}, \citenamefont {Mui{\~{n}}os-Landin},
  \citenamefont {Fischer}, \citenamefont {Söker}, \citenamefont {Cichos},
  \citenamefont {Kapral}, \citenamefont {Gaspard}, \citenamefont {Ripoll},
  \citenamefont {Sagues}, \citenamefont {Doostmohammadi}, \citenamefont
  {Yeomans}, \citenamefont {Aranson}, \citenamefont {Bechinger}, \citenamefont
  {Stark}, \citenamefont {Hemelrijk}, \citenamefont {Nedelec}, \citenamefont
  {Sarkar}, \citenamefont {Aryaksama}, \citenamefont {Lacroix}, \citenamefont
  {Duclos}, \citenamefont {Yashunsky}, \citenamefont {Silberzan}, \citenamefont
  {Arroyo},\ and\ \citenamefont {Kale}}]{Gompper_2020}%
  \BibitemOpen
  \bibfield  {author} {\bibinfo {author} {\bibfnamefont {G.}~\bibnamefont
  {Gompper}}, \bibinfo {author} {\bibfnamefont {R.~G.}\ \bibnamefont
  {Winkler}}, \bibinfo {author} {\bibfnamefont {T.}~\bibnamefont {Speck}},
  \bibinfo {author} {\bibfnamefont {A.}~\bibnamefont {Solon}}, \bibinfo
  {author} {\bibfnamefont {C.}~\bibnamefont {Nardini}}, \bibinfo {author}
  {\bibfnamefont {F.}~\bibnamefont {Peruani}}, \bibinfo {author} {\bibfnamefont
  {H.}~\bibnamefont {Löwen}}, \bibinfo {author} {\bibfnamefont
  {R.}~\bibnamefont {Golestanian}}, \bibinfo {author} {\bibfnamefont {U.~B.}\
  \bibnamefont {Kaupp}}, \bibinfo {author} {\bibfnamefont {L.}~\bibnamefont
  {Alvarez}}, \bibinfo {author} {\bibfnamefont {T.}~\bibnamefont {Ki{\o}rboe}},
  \bibinfo {author} {\bibfnamefont {E.}~\bibnamefont {Lauga}}, \bibinfo
  {author} {\bibfnamefont {W.~C.~K.}\ \bibnamefont {Poon}}, \bibinfo {author}
  {\bibfnamefont {A.}~\bibnamefont {DeSimone}}, \bibinfo {author}
  {\bibfnamefont {S.}~\bibnamefont {Mui{\~{n}}os-Landin}}, \bibinfo {author}
  {\bibfnamefont {A.}~\bibnamefont {Fischer}}, \bibinfo {author} {\bibfnamefont
  {N.~A.}\ \bibnamefont {Söker}}, \bibinfo {author} {\bibfnamefont
  {F.}~\bibnamefont {Cichos}}, \bibinfo {author} {\bibfnamefont
  {R.}~\bibnamefont {Kapral}}, \bibinfo {author} {\bibfnamefont
  {P.}~\bibnamefont {Gaspard}}, \bibinfo {author} {\bibfnamefont
  {M.}~\bibnamefont {Ripoll}}, \bibinfo {author} {\bibfnamefont
  {F.}~\bibnamefont {Sagues}}, \bibinfo {author} {\bibfnamefont
  {A.}~\bibnamefont {Doostmohammadi}}, \bibinfo {author} {\bibfnamefont
  {J.~M.}\ \bibnamefont {Yeomans}}, \bibinfo {author} {\bibfnamefont {I.~S.}\
  \bibnamefont {Aranson}}, \bibinfo {author} {\bibfnamefont {C.}~\bibnamefont
  {Bechinger}}, \bibinfo {author} {\bibfnamefont {H.}~\bibnamefont {Stark}},
  \bibinfo {author} {\bibfnamefont {C.~K.}\ \bibnamefont {Hemelrijk}}, \bibinfo
  {author} {\bibfnamefont {F.~J.}\ \bibnamefont {Nedelec}}, \bibinfo {author}
  {\bibfnamefont {T.}~\bibnamefont {Sarkar}}, \bibinfo {author} {\bibfnamefont
  {T.}~\bibnamefont {Aryaksama}}, \bibinfo {author} {\bibfnamefont
  {M.}~\bibnamefont {Lacroix}}, \bibinfo {author} {\bibfnamefont
  {G.}~\bibnamefont {Duclos}}, \bibinfo {author} {\bibfnamefont
  {V.}~\bibnamefont {Yashunsky}}, \bibinfo {author} {\bibfnamefont
  {P.}~\bibnamefont {Silberzan}}, \bibinfo {author} {\bibfnamefont
  {M.}~\bibnamefont {Arroyo}}, \ and\ \bibinfo {author} {\bibfnamefont
  {S.}~\bibnamefont {Kale}},\ }\href {\doibase 10.1088/1361-648x/ab6348}
  {\bibfield  {journal} {\bibinfo  {journal} {Journal of Physics: Condensed
  Matter}\ }\textbf {\bibinfo {volume} {32}},\ \bibinfo {pages} {193001}
  (\bibinfo {year} {2020})}\BibitemShut {NoStop}%
\bibitem [{\citenamefont {Vicsek}\ \emph
  {et~al.}(1995{\natexlab{b}})\citenamefont {Vicsek}, \citenamefont
  {Czir{\'{o}}k}, \citenamefont {Ben-Jacob}, \citenamefont {Cohen},\ and\
  \citenamefont {Shochet}}]{Vicsek1995}%
  \BibitemOpen
  \bibfield  {author} {\bibinfo {author} {\bibfnamefont {T.}~\bibnamefont
  {Vicsek}}, \bibinfo {author} {\bibfnamefont {A.}~\bibnamefont
  {Czir{\'{o}}k}}, \bibinfo {author} {\bibfnamefont {E.}~\bibnamefont
  {Ben-Jacob}}, \bibinfo {author} {\bibfnamefont {I.}~\bibnamefont {Cohen}}, \
  and\ \bibinfo {author} {\bibfnamefont {O.}~\bibnamefont {Shochet}},\ }\href
  {http://link.aps.org/doi/10.1103/PhysRevLett.75.1226} {\bibfield  {journal}
  {\bibinfo  {journal} {Phys. Rev. Lett.}\ }\textbf {\bibinfo {volume} {75}},\
  \bibinfo {pages} {1226} (\bibinfo {year} {1995}{\natexlab{b}})}\BibitemShut
  {NoStop}%
\bibitem [{\citenamefont {Chat{\'{e}}}\ \emph {et~al.}(2008)\citenamefont
  {Chat{\'{e}}}, \citenamefont {Ginelli},\ and\ \citenamefont
  {Raynaud}}]{Chate2008}%
  \BibitemOpen
  \bibfield  {author} {\bibinfo {author} {\bibfnamefont {H.}~\bibnamefont
  {Chat{\'{e}}}}, \bibinfo {author} {\bibfnamefont {F.}~\bibnamefont
  {Ginelli}}, \ and\ \bibinfo {author} {\bibfnamefont {F.}~\bibnamefont
  {Raynaud}},\ }\href {\doibase 10.1103/PhysRevE.77.046113} {\bibfield
  {journal} {\bibinfo  {journal} {Phys. Rev. E}\ }\textbf {\bibinfo {volume}
  {77}},\ \bibinfo {pages} {046113} (\bibinfo {year} {2008})}\BibitemShut
  {NoStop}%
\bibitem [{Note1()}]{Note1}%
  \BibitemOpen
  \bibinfo {note} {It is not difficult to generalise the model to do
  so}\BibitemShut {NoStop}%
\bibitem [{\citenamefont {Kampen}(2001)}]{vankampen}%
  \BibitemOpen
  \bibfield  {author} {\bibinfo {author} {\bibfnamefont {N.~G.~V.}\
  \bibnamefont {Kampen}},\ }\href@noop {} {\emph {\bibinfo {title} {{Stochastic
  processes in physics and chemistry}}}}\ (\bibinfo  {publisher} {Elsevier},\
  \bibinfo {address} {Amsterdam},\ \bibinfo {year} {2001})\BibitemShut
  {NoStop}%
\bibitem [{\citenamefont {Risken}(1984)}]{Risken}%
  \BibitemOpen
  \bibfield  {author} {\bibinfo {author} {\bibfnamefont {H.}~\bibnamefont
  {Risken}},\ }\href@noop {} {\emph {\bibinfo {title} {{The Fokker-Planck
  Equation}}}}\ (\bibinfo  {publisher} {Springer: Berlin},\ \bibinfo {year}
  {1984})\BibitemShut {NoStop}%
\bibitem [{Note2()}]{Note2}%
  \BibitemOpen
  \bibinfo {note} {For modelling the dynamics of physical problems at
  equilibrium, one often chooses the discretisation that gives rise to the
  correct steady state equilibrium probability distribution.}\BibitemShut
  {Stop}%
\bibitem [{\citenamefont {Markowich}\ and\ \citenamefont
  {Villani}(2000)}]{Markowich2000}%
  \BibitemOpen
  \bibfield  {author} {\bibinfo {author} {\bibfnamefont {P.~A.}\ \bibnamefont
  {Markowich}}\ and\ \bibinfo {author} {\bibfnamefont {C.}~\bibnamefont
  {Villani}},\ }\href {\doibase 10.1.1.35.2278} {\bibfield  {journal} {\bibinfo
   {journal} {Mat. Contemp.}\ }\textbf {\bibinfo {volume} {19}},\ \bibinfo
  {pages} {1} (\bibinfo {year} {2000})}\BibitemShut {NoStop}%
\bibitem [{\citenamefont {Guionnet}\ and\ \citenamefont
  {Zegarlinski}(2002)}]{Guionnet2002}%
  \BibitemOpen
  \bibfield  {author} {\bibinfo {author} {\bibfnamefont {A.}~\bibnamefont
  {Guionnet}}\ and\ \bibinfo {author} {\bibfnamefont {B.}~\bibnamefont
  {Zegarlinski}},\ }\href
  {http://www.numdam.org/item/SPS{\_}2002{\_}{\_}36{\_}{\_}1{\_}0/} {\bibfield
  {journal} {\bibinfo  {journal} {S{\'{e}}minaire Probab. Strasbg.}\ }\textbf
  {\bibinfo {volume} {36}},\ \bibinfo {pages} {1} (\bibinfo {year}
  {2002})}\BibitemShut {NoStop}%
\bibitem [{\citenamefont {Liverpool}(2020)}]{Liverpool2020}%
  \BibitemOpen
  \bibfield  {author} {\bibinfo {author} {\bibfnamefont {T.~B.}\ \bibnamefont
  {Liverpool}},\ }\href {\doibase 10.1103/PhysRevE.101.042107} {\bibfield
  {journal} {\bibinfo  {journal} {Phys. Rev. E}\ }\textbf {\bibinfo {volume}
  {101}},\ \bibinfo {pages} {042107} (\bibinfo {year} {2020})}\BibitemShut
  {NoStop}%
\bibitem [{\citenamefont {Cameron}\ \emph {et~al.}(2023)\citenamefont
  {Cameron}, \citenamefont {Mosayebi}, \citenamefont {Bennett},\ and\
  \citenamefont {Liverpool}}]{Cameron2023}%
  \BibitemOpen
  \bibfield  {author} {\bibinfo {author} {\bibfnamefont {S.}~\bibnamefont
  {Cameron}}, \bibinfo {author} {\bibfnamefont {M.}~\bibnamefont {Mosayebi}},
  \bibinfo {author} {\bibfnamefont {R.}~\bibnamefont {Bennett}}, \ and\
  \bibinfo {author} {\bibfnamefont {T.~B.}\ \bibnamefont {Liverpool}},\ }\href
  {\doibase 10.1103/PhysRevE.108.014608} {\bibfield  {journal} {\bibinfo
  {journal} {Phys. Rev. E}\ }\textbf {\bibinfo {volume} {108}},\ \bibinfo
  {pages} {1} (\bibinfo {year} {2023})},\ \Eprint
  {http://arxiv.org/abs/2201.10813} {arXiv:2201.10813} \BibitemShut {NoStop}%
\bibitem [{\citenamefont {Higham}(2001)}]{Higham2001}%
  \BibitemOpen
  \bibfield  {author} {\bibinfo {author} {\bibfnamefont {D.~J.}\ \bibnamefont
  {Higham}},\ }\href {\doibase 10.1137/S0036144500378302} {\bibfield  {journal}
  {\bibinfo  {journal} {SIAM Rev.}\ }\textbf {\bibinfo {volume} {43}},\
  \bibinfo {pages} {525} (\bibinfo {year} {2001})}\BibitemShut {NoStop}%
\bibitem [{\citenamefont {Derrida}(2007)}]{Derrida2007}%
  \BibitemOpen
  \bibfield  {author} {\bibinfo {author} {\bibfnamefont {B.}~\bibnamefont
  {Derrida}},\ }\href {\doibase 10.1088/1742-5468/2007/07/P07023} {\bibfield
  {journal} {\bibinfo  {journal} {J. Stat. Mech. Theory Exp.}\ } (\bibinfo
  {year} {2007}),\ 10.1088/1742-5468/2007/07/P07023},\ \Eprint
  {http://arxiv.org/abs/0703762} {arXiv:0703762 [cond-mat]} \BibitemShut
  {NoStop}%
\bibitem [{\citenamefont {Jona-Lasinio}(2023)}]{Jona-Lasinio2023}%
  \BibitemOpen
  \bibfield  {author} {\bibinfo {author} {\bibfnamefont {G.}~\bibnamefont
  {Jona-Lasinio}},\ }\href {\doibase 10.5194/npg-30-253-2023} {\bibfield
  {journal} {\bibinfo  {journal} {Nonlinear Process. Geophys.}\ }\textbf
  {\bibinfo {volume} {30}},\ \bibinfo {pages} {253} (\bibinfo {year}
  {2023})}\BibitemShut {NoStop}%
\end{thebibliography}%

\end{document}